\documentclass[numberedappendix]{emulateapj}
\usepackage{apjfonts}

\newcommand{\timav}{\langle \dot M \rangle}

\shorttitle{The Thermal state of WDs undergoing CN}

\begin{document}

\submitted{Accepted to the ApJ}
\title{Theoretical modeling of the thermal state of accreting white
dwarfs undergoing classical novae}

\author{Dean M. Townsley}
\affil{Department of Physics\\
Broida Hall, University of California, Santa Barbara, CA 93106;
townsley@physics.ucsb.edu}
\author{Lars Bildsten}
\affil{Kavli Institute for Theoretical Physics and Department of Physics\\
Kohn Hall, University of California, Santa Barbara, CA 93106;
bildsten@kitp.ucsb.edu}

\begin{abstract}
 
White dwarfs experience a thermal renaissance when they receive mass
from a stellar companion in a binary. For accretion rates $<10^{-8}
M_\odot \ {\rm yr^{-1}}$, the freshly accumulated hydrogen/helium
envelope ignites in a thermally unstable manner that results in a
classical novae (CN) outburst and ejection of material. We have
undertaken a theoretical study of the impact of the accumulating
envelope on the thermal state of the underlying white dwarf (WD). This
has allowed us to find the equilibrium WD core temperatures ($T_c$), the
classical nova ignition masses ($M_{\rm ign}$) and the thermal luminosities for WDs
accreting at rates of $10^{-11}-10^{-8} M_\odot \ {\rm yr^{-1}}$.
These accretion rates are most appropriate to WDs in cataclysmic
variables (CVs) of $P_{\rm orb}\lesssim 7$ hr, many of which accrete
sporadically as dwarf novae. We have included $^3$He in the accreted
material at levels appropriate for CVs and find that it
significantly modifies the CN ignition mass.
We compare our results with several others from the
CN literature and find that the inclusion of
$^3$He leads to lower $M_{\rm ign}$ for $\timav \gtrsim
10^{-10}M_\odot\ {\rm yr^{-1}}$, and that for $\timav$ below this the
particular author's assumption concerning $T_c$, which we calculate
consistently, is a determining factor.
Initial
comparisons of our CN ignition masses with measured ejected masses
find reasonable agreement and point to ejection of material
comparable to that accreted. 

\end{abstract}



\section{Introduction}

 Classical novae (CN) are the result of unstable thermonuclear
ignition of freshly accreted hydrogen and helium on a white dwarf (WD)
in a mass transferring binary. The orbital periods of these binaries
presently range from 1.4 to more than 16 hours, with a large number in
the 2-4 hour range \citep{DiazBruc97,Warn02}. These cataclysmic
variables (CVs; \citealt{Warn95}) have low-mass stellar companions and
spend most of their lives accreting at time averaged rates,
$\timav\approx 10^{-9}-10^{-11} M_\odot \ {\rm yr^{-1}}$ (e.g.\
Howell, Nelson, \& Rappaport 2001). The frequency of CN as a function
of orbital period depends on the accumulated mass immediately prior to
the explosion, $M_{\rm ign}$. However, the value of $M_{\rm ign}$
strongly depends on the WD's core temperature, $T_c$, which is {\it a
priori} unknown and must either be calculated or measured. Knowledge
of the ignition mass is also needed for comparisons to measured
ejected masses in CN that would decide whether or not the underlying
carbon/oxygen WD is being excavated during the CN.

 CVs are formed when the WD made during a common envelope event
finally comes into contact with its companion as a result of
gravitational radiation losses over a few Gyr (e.g.~Howell et al.\ 2001). The WD
will have cooled during this time; a $0.20 M_\odot$ He WD would have
$T_c=3.3\times 10^6 \ {\rm K} $ at 4 Gyr \citep{AlthBenv97}, whereas a
$0.6M_\odot$ C/O WD would have $T_c=2.5\times 10^6\ {\rm K}$ in 4 Gyr
\citep{Salaetal00}. These WDs would have effective temperatures $T_{\rm
eff}\approx 4500-5000$ K. A subset of the CVs, called dwarf novae
(DN), contain a WD accreting at low time-averaged rates
$\timav<10^{-9}M_\odot \ {\rm yr}^{-1}$, where the accretion disk is
subject to a thermal instability which causes it to rapidly transfer
matter onto the WD (at $\dot M \gg \timav$) for roughly a week once every
month to year. The $\dot M$ onto the WD is low enough between
outbursts that the UV emission is dominated by the internal luminosity
of the WD, allowing for a measurement of
$T_{\rm eff}>10,000\ {\rm K}$ \cite{Sion99}.
This is the best evidence
that the WD is hotter than a cooling WD of similar age
due to the thermal impact of accretion.
\citet{Sion95} has argued that most of this heating is from
gravitational energy released (i.e. converted to thermal energy) in
the WD interior via material compression.  This energy is then
transported outward to the stellar surface, and we find that it can
similarly heat the core.
We will discuss this in great detail, as
well as explain the important role of nuclear ``simmering''. 
A thorough comparison of our theoretical work to $T_{\rm eff}$
observations and a summary of the implications for CV evolution are in
a companion paper \citep{TownBild03}.

The gravitational energy released ($GM/R$) when a particle falls from
a large distance to the stellar surface, sometimes termed the
accretion energy, is deposited at, or near, the
photosphere and is rapidly radiated away. This energy does not get
taken into the star, as the time it takes the fluid to move inward is
always much longer than the time it takes for heat to escape. This
basically eliminates the outer boundary condition and instead points
to the importance of energy released deep in the WD due to both
gravitational energy released, via compression, within the star
and hydrostatic nuclear
burning. That energy takes a long time to exit, will still be visible
when accretion has halted, sets $T_{\rm eff}$, and reheats the WD.

 We begin in \S \ref{sec:accenv} by explaining the physics of
heating within the accreted envelope and how that
differs from heating in the deep core (discussed in Appendix
A). Since these WDs are accreting at rates low enough so that the
accumulated hydrogen burns unstably, we also need to carefully track
the accumulated mass of material and account for the additional energy
source of slow nuclear ``simmering'' near the base of the accreted
layer. This is significant as the layer accumulates and eventually
triggers the CN. The calculation of $M_{\rm ign}$ for a fixed $T_c$ is 
explained in \S \ref{sec:ign} (closely comparing to the original
Fujimoto 1982 work), where we also exhibit the importance of
the $^3{\rm He}$ content in the accreting material. 

Our method of finding the equilibrium WD core temperature (Townsley \& Bildsten
2002) is explained in \S \ref{sec:equil} and improves on that of
\citet{Ibenetal92a} by allowing the accreted envelope mass to change
through the CN cycle. We show that early in the cycle, when the mass
of the newly accreted layer is small, compressional heating is small,
and the WD cools. Later in the cycle, the accreted layer becomes thick
enough that compressional heating along with nuclear simmering heats
the core.  So, the core cools at low accumulated masses and is heated
prior to unstable ignition.  Using CN ignition conditions to determine the
maximum mass of the overlying freshly accreted shell, we then find the
steady-state (i.e.  cooling equals heating throughout the CN cycle)
core temperature, $T_{\rm c,eq}$, as a function of $\timav$ and $M$.
We close \S \ref{sec:equil} by showing that for $\timav \lesssim
10^{-9}M_\odot$ yr$^{-1}$ the WD reaches this
equilibrium after accreting an amount of mass less than its own
and that in CVs with $P_{\rm orb}\lesssim2$ hours the accretion rate
changes slowly enough that the WD should always  be near the
equilibrium given by $\timav$ and $M$. 

 Having found the core temperatures, we then predict the amount of
mass that needs to accumulate on the WD in order to unstably ignite
the accumulated hydrogen as a function of $\timav$ and $M$. Our work
is compared to previous calculations in \S \ref{sec:res}, finding
very large differences with others due to their lack of knowledge of
$T_c$.  We also compare our predictions of $M_{\rm ign}$ to those few
measures of the ejected masses in CN of known orbital period. Within
the confines of this limited comparison, we find that the ejected
masses are comparable to the amount accumulated.
The equilibrium also yields powerful relations between the WD surface
temperature and its accretion rate and mass.  These allow for tests of
binary evolution models as well as measures of the masses of these
WDs; a comparison to the accumulated observations of effective
temperatures in the DN systems is being published separately
\citep{TownBild03}. We close with a summary and discussion of future
work, especially the seismology of these accreting WDs.

\section{Atmospheric model including accretion}
\label{sec:accenv}

Starting with \citet{Mest52} single WD atmospheres are characterized by a
relation between the surface luminosity, $L$, and the core
temperature, $T_c$ \citep[e.g.][]{Wood95,Hans99,Chabetal00,Salaetal00}.  This
is one of two components (the other being the thermal content of the
core) needed for evolutionary calculations of a cooling WD.  In this
case the atmosphere extends to the location where electron conduction
forces the core to be isothermal. This occurs at $M_{\rm
env}=5\times 10^{-5}M_\odot$ for $T_c=4\times10^6$ K and
$M=0.6M_\odot$ when $L=3\times 10^{-4}L_\odot$ (see Figure 1 in
Fontaine, Brassard, \& Bergeron 2001).

 In an accreting WD, energy is released in the atmosphere by the slow
downward motion of matter in the gravitational field and by a slow
rate of hydrogen fusion. This internal energy release modifies the
$L$-$T_c$ relation, giving an accreting WD a higher $L$ than a
non-accretor of the same $T_c$.  The $L$-$T_c$ relation of
non-accreting WDs depends on the surface gravity, $g$, which is set by
the mass, $M$, and on the composition of the envelope (which fixes the
opacity), typically split into layers of pure H and He. The WD mass
enters in much the same way in the accreting case, but the composition
also determines the local energy release in the envelope. We
parameterize composition by the mass of the freshly accreted layer,
$M_{\rm acc}$, which has the composition of the donor star. 
We use a solar composition with mass fractions:
$X_{\rm H}=0.709$, $X_{^4\rm He}=0.271$, $X_{\rm C}=0.009$, and
$X_{\rm O}=0.011$.  Since $^3{\rm He}$ will prove important as an
ignition trigger (see discussion in \S \ref{sec:ign}) at
low $\timav$, we will show
models with $X_{^3\rm He}=0.001$ and $0.005$, the lower and upper limits on expected
values for the accreted material (e.g.~\citealt{DAntMazz82}; \citealt{IbenTutu84}).
Interior to this layer
we assume a C/O core.
As material is accreted at a time averaged rate $\timav$, $M_{\rm
acc}$ varies from zero (immediately after a classical nova) to some
maximum, $M_{\rm ign}$, just before ignition of the next classical
nova.  Thus three parameters, $M$, $M_{\rm acc}$ and $\timav$ lead to
an $L$-$T_c$ relation.

Due to the deposition of energy in the envelope, the surface
luminosity is not equal to the thermal energy lost by the degenerate
core.  However, the atmospheric model provides not only $L$ at the
surface, but also the luminosity, $L_{\rm core}$, which enters the
bottom of the atmosphere from the core.  This luminosity couples the
envelope model to the thermal content of the core, determining the
temporal evolution of $T_c$.  As we will show, when $T_c$ is low, 
the envelope heats the degenerate core and $L_{\rm core}<0$.

\subsection{Thin Shell Approximation}

As a spherical, gravitationally self-bound object accretes matter, it
undergoes structural changes which we divide into two types: (1)
global structural changes which affect the core radius, and (2)
internal movements of material.  See \citet{Nomo82} and Appendix A
for a discussion of the distinction between these terms. In the outer parts
of the WD, the two effects are largely decoupled; material moves
through the envelope and down into a deeper reservoir and the WD
radius and the gravitional field in which the envelope exists
are set by the total WD mass.
This is particularly useful for
the case of low $\timav$, where the global changes happen extremely
slowly, and can therefore be safely neglected. Our treatment of the
core is presented in Appendix A, where we show that the energy release
in the envelope dominates that in the core by a factor $\simeq 5$.

 Within a thin shell near the surface, accreted material enters at the
top, and displaces pre-existing material which moves 'out' of the
shell, deeper into the star.  With the assumption that the radius and
total stellar mass, and thus $g$, are fixed on the timescale being
considered, the local heat equation becomes
\begin{equation}
\label{eq:heateq}
T\frac{Ds}{Dt}=T\frac{\partial s}{\partial
t}+Tv_r \frac{\partial s}{\partial r}= -\frac{d L}{d M_r} +\epsilon_N,
\end{equation}
where $\epsilon_N$ is the nuclear burning rate, $s$ is the entropy per
gram, and $v_r=-\timav /4\pi r^2 \rho$ is the downward speed of
material motion in response to accretion.  If a fluid element loses
entropy as it is compressed, then that liberated energy
is carried away by heat transport.  The thermal structure of
the envelope determines the two spatial derivatives and the nuclear
energy input in equation (\ref{eq:heateq}).  Thus $\partial s/\partial t$,
the time evolution  of the entropy field, depends on the current
thermal structure and $\timav$.  A thermal structure for which
$\partial s/\partial t=0$ is a static state for the envelope.  If a
higher $\timav$ were applied without changing the thermal structure, a
local buildup of heat would occur, $\partial s/\partial t>0$, due to
the increase of $|v_r|$ without a change in $dL/dM_r$.  A lower
$\timav$ leads to a corresponding cooling of the envelope.  When the
entropy advection term dominates over the nuclear burning term,
this static state is stable because $L$ has a stronger temperature
dependence than $s$ in both degenerate and nondegenerate regimes.
However, when $\epsilon_N$  dominates over the advection a thermal
instability is likely, leading to a classical nova explosion.

The radial derivative is evaluated from the
envelope structure.  We work in pressure coordinates, and
define $\dot m=\timav/4\pi r^2$, so
\begin{equation}
\label{eq:Tvdsdt}
Tv_r\frac{d s}{d r}=
g\dot m T\frac{ds}{dP}
=\frac{g\dot m c_PT}{P}\left[\frac{d \ln T}{d \ln P} -
\nabla_{\rm ad}\right]\ ,
\end{equation}
where $c_P=T(\partial s/\partial T)_P$ is the specific heat at constant
pressure and $\nabla_{\rm ad}=(\partial\ln T/\partial\ln P)_s$.
The entropy advection is given by how much the temperature gradient
differs from the adiabat. In the static state, the
gradient of the luminosity itself is then
\begin{equation}
\label{dLdP}
\frac{d L}{dP}=\timav c_P
\left[\frac{d T}{d P} -
\left(\frac{\partial T}{\partial P}\right)_s\right]
-\frac{4\pi r^2\epsilon_N}{g}\ ,
\end{equation}
which we use with the equations for heat transport and hydrostatic
equilibrium to solve for both the structure and luminosity in the
envelope, resulting in an $L$ and $L_{\rm core}$ which depend on
$T_c$, $M$, $\timav$, and $M_{\rm acc}$.  The contribution
to $L$ from accretion is the heat not advected with the
material, and is
referred to in the literature as ``compressional heating.''  Though
this choice of words is unfortunate since the action occurring is more
like cooling, we use this common term in our discussions.
In all the calculations presented here we use a fixed $R$ for the last
term in equation (\ref{dLdP}),  that of a zero temperature WD, $8.8$,
$5.7$ and $4.1\times 10^8$ cm for $M=0.6$, $1.0$ and $1.2M_\odot$
respectively.  As a single point check, matching the bottom
of the envelope and the outer edge of a finite-temperature core model
yeilds $R=8.9\times10^8$ cm for $M=0.6M_\odot$, $\timav = 6\times
10^{-11}M_\odot\ {\rm yr^{-1}}$, and $M_{\rm acc}=1.4\times
10^{-4}M_\odot$.
By dropping the nuclear burning term and making a few
other simplifying assumptions equation (\ref{dLdP}) can be integrated
analytically to obtain an approximate luminosity, $L\approx 3 \timav
kT_c/\mu m_p$ (see Appendix \ref{app:estimate}),  where $k$ is
Boltzmann's constant, $m_p$ is the proton mass and $\mu\simeq 0.6$ is
the mean molecular weight of the freshly accreted envelope. 

Since $M_{\rm acc}$ changes with time as $M_{\rm acc}(t)=\timav \Delta t$, 
the static envelope structure is an approximation to the 
evolution. For this method to be
accurate, we want the neglected term to be small, namely
$|\partial s/\partial t| \ll |v_r\ \partial s/\partial r|$.
Evaluating at a fixed pressure and using equation (\ref{eq:Tvdsdt}) gives
\begin{equation}
\left|\frac{\partial T}{\partial t}\right| \ll
\frac {T}{\tau_{\rm acc}}\left|\nabla_{\rm ad}-\frac{d\ln T}{d\ln P}\right|
\ ,
\end{equation}
where $\tau_{\rm acc}=P/g\dot m$ is the accretion time.  Dropping the
temperature gradient (good in the conductive regions), finite differencing
and setting $\Delta t = \tau_{\rm acc}$ leads to the condition $\Delta T/T\ll
2/5$, where $\Delta T$ is the temperature change at the envelope base during
accumulation.  As will be shown in Section \ref{sec:equil}, the evolution of
$M_{\rm acc}$ takes place at constant $T_c$, so that this condition is always
satisfied and our static structures are an excellent approximation for
the accumulating envelope. 

Now we wish to consider the effect of the outer boundary condition on the
structure of the envelope.  To determine this we compare
the thermal cooling time of an outer layer, $\tau_{\rm th}=\Delta Mc_P T/L$
to the time to accrete that layer $\tau_{\rm acc}=\Delta M/\timav$. Here
$\Delta M$ is the mass of the (thin) outer layer being considered, $T$ is the
temperature of the layer, and $L$ is the overall luminosity which is driven
mostly by energy released from deeper layers, $L\approx c_p\timav T_c$.  Thus
the ratio $\tau_{\rm th}/\tau_{\rm acc} \approx T/T_c$, so that where $T\ll
T_c$ in the outer layers, the material can cool much more quickly than it is
being accreted and therefore has no influence over the thermal structure of
deeper layers.  In other words, the entropy of the outer envelope is
determined by the luminosity flowing through it, rather than any external
boundary condition.  The condition $T\ll T_c$ is sometimes violated during a
dwarf novae (DN) event, where either the high external radiation field
\citep{Prin88} or the more rapid accretion \citep{Sion95,GodoSion02}
temporarily modifies the outermost layers of the WD envelope.  However, this
heat does not penetrate to depths critical for the WD core, but only to a
depth where the thermal time is of order the duration of the DN outburst, and
is the likely cause of the fading $T_{\rm eff}$ seen following a DN outburst
(see Sion's 1999 review).

In addition to the downward motion due to accretion, the heavier
Helium ions will sink with respect to the overall flow.  While we have
not included this effect, we estimate here when it is important.  The
drift velocity of the He ions is $v_{\rm drift} \simeq 2m_p gD/kT$
where $D$ is the thermal diffusion coefficient (See
\citealt{DeloBild02}).  This is to be compared with the downward
velocity of the fluid due to accretion, $v_{\rm acc} = \timav/4\pi
R^2\rho$.  In the simplest approximation, Coulomb scattering off
protons dominates and $D = 0.024\Lambda^{-1}T_7^{5/2}\rho_4^{-1}$
cm$^2$ s$^{-1}$ \citep{AlcoIlla80}, where $T_7= T/10^7$ K and $\rho_4
= \rho/10^4$ g cm$^{-3}$ and $\Lambda$ is the Coulomb logarithm.  For
$v_{\rm drift}< v_{\rm acc}$ this yields the relationship $\timav >
10^{-11}M_\odot\ {\rm yr^{-1}} T_7^{3/2}(M/0.6M_\odot)\Lambda^{-1}$,
so that for our lowest accretion rate, $10^{-11}M_\odot$ yr$^{-1}$,
where $T_7= 0.4$, only the high mass models deserve further scrutiny.
Below this accretion rate, the energy release due to sedimentation of
the Helium in the accreted material can be a moderate fraction of that
released due to compression.  For our $M=1.0M_\odot$ model with
$\timav = 10^{-11}M_\odot$ yr$^{-1}$, the conditions at the base of
the H/He layer when $M_{\rm acc} = 0.90 M_{\rm ign}$ are
$\rho=1.4\times 10^4$ g cm$^{-3}$, $T=4.1\times 10^6$ K so that the
Coulomb coupling parameter $\Gamma= 2$.  Diffusion coefficients in
this gas-liquid crossover regime have yet to be authoritatively
calculated.  Using the fitting form of \citet{IbenMacD85}, which is
very similar to the canonical results of \citet{Paqu86} in this regime
(See \citealt{Ibenetal92b}), we obtain $v_{\rm drift}/v_{\rm acc} =
1.3$.  However, using the liquid diffusion results of
\citet{WallBaus78}, we obtain $v_{\rm drift}/v_{\rm acc} = 0.38$.
Regardless of this uncertainty, it is clear that in future work
diffusion needs to be considered for $M\ge 1.0M_\odot$ and $\timav \le
10^{-11}M_\odot$ yr$^{-1}$.  As inferred from the amount of
sedimentation found by \citet{Ibenetal92b}, this
limit is at higher $\timav$ when a higher $T_c$ is assumed.

\subsection{Accreting Envelope}
\label{sec:ourenv}

We now present results from the envelope models, investigating how
$\timav$ and  $M_{\rm acc}$ determine the $L$-$T_c$ relationship.  The
outer boundary is at $P=10^{10}$ erg cm$^{-3}$, in the upper edge of
the radiative layers.  We set the temperature there to that of a
radiative zero solution for $L= 4\pi R^2\sigma_{\rm SB}T_{\rm eff}^4$.
Then equation (\ref{dLdP}) along with the other structure equations are
integrated inward, keeping track of $L(P)$ and $T(P)$, to $P= gM_{\rm
acc}/4\pi R^2$, the base of the accreted layer.  Interior to this,
$L=L_{\rm core}$ is taken to be constant and the integration is
continued, using a pure carbon composition for simplicity, until $P=
40 P_{\rm ign}$, where $P_{\rm ign} = g M_{\rm ign}/4\pi R^2$ and
$M_{\rm ign}$ is found consistently at each $\timav$.  A
value of $P=2\times 10^{20}$ erg cm$^{-3}$ is used for the
$M=0.6M_\odot$ demonstrations in this section, corresponding to
the $M_{\rm ign}$ for an equilibrated model at $\timav =
10^{-10}M_\odot$ yr$^{-1}$.  We label the temperature at the point
where we stop the integration $T_c$, interior to which the
core is isothermal.  The equation of state for the electrons and ions
was taken from analytical approximations of \citet{Pacz83} with
Coulomb corrections from \citet{FaroHama93}.  The radiative opacities
from OPAL \citep{IgleRoge96} and the conductivities of
\citet{Itohetal83} were used.  Nuclear reaction rates and energies are
from \citet{CaugFowl88} with screening enhancement factors for high density from
\citet{Ichi93}, which also appear in \citet{Ogataetal91}

The above treatment of the outer edge of the C/O core provides a
reasonable approximation to the evolution of this region during a CN
cycle without the complications of explicit simulation.  At $40 P_{\rm
ign}$ the temperature change due to adiabatic compression is 1\%.
Under this approximation in the equilibrium state discussed in
\S\ref{sec:equil}, fluid elements near the base of the accreted layer
move on paths for which they must lose entropy during the CN cycle,
i.e. ones more shallow than the adiabat in the $T$-$P$ plane. Deeper
fluid elements move on paths for which they must gain entropy, steeper
than the adiabat.  This exchange is consistent with our expectation
for the actual evolution, and since it only contributes peripherally
by setting the gross temperature excursion from $T_c$ at the base of
the accreted layer, the accuracy of a more sophisticated treatment is
not considered worth its price in complexity at this point.  The
fraction of the core affected by the CN cycle in this treatment is
also roughly consistent with that expected on energetic grounds.  For
example, a luminosity of $10^{-3}L_\odot$ delivers enough energy in
the interval between classical novae, $\sim10^6$ years, to raise the
temperature of the outer $10^{-3}M_\odot$ of the core by $10^6$ K
($\sim10\%$), this mass coordinate corresponds to a pressure of just
under $10^{20}$ erg cm$^{-3}$.

\begin{figure}
\plotone{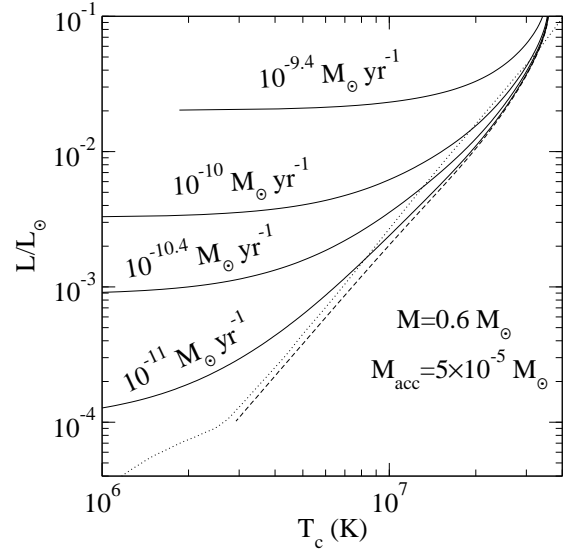}
\caption{
\label{fig:L-Tc}
Exiting surface luminosity, $L$, as a function of core temperature,
$T_c$, for a $0.6M_\odot$ WD.  The accumulated H/He layer has a mass
$M_{\rm acc}=5\times10^{-5}M_\odot$.  Various accretion rates are
shown as labelled solid lines with $\timav=10^{-11}$, $10^{-10.4}$,
$10^{-10}$, and $10^{-9.4}M_\odot$ yr$^{-1}$.  The dashed line has
$\timav=0$, and the dotted line is the $L$-$T_c$ relation of an
isolated cooling DA WD with pure layers of mass $M_{\rm
H}=6\times10^{-5}M_\odot$ and $M_{\rm He}=6\times10^{-3}M_\odot$
\citep{Hans99}.}
\end{figure}

The relationship between $L$ and $T_c$ for several $\timav$'s are
shown in Figure \ref{fig:L-Tc} for a $0.6 M_\odot$ WD ($R=8.76\times
10^8$ cm) with $M_{\rm acc}=5\times 10^{-5}M_\odot$. This $M_{\rm
acc}$ is less than the CN ignition mass for most of the range of $\timav$
and is comparable to the hydrogen
layers on cooling WDs.  All the demonstration curves in this section
are for $X_{^3\rm He} =0.001$.  Shown for reference (dotted line)
is the $L$-$T_c$ relation for an isolated $0.6M_\odot$ WD with pure
layers of mass $M_{\rm H}=6\times10^{-5}M_\odot$ and $M_{\rm
He}=6\times10^{-3}M_\odot$ \citep{Hans99}. The corresponding relation
for our envelope model, with $\timav=0$, is shown as the dashed line
and provides an initial comparison to previous work. The offset from
the cooling WD is due to the H/He envelope being more opaque than that
of pure hydrogen.
The slightly larger divergence for $10^7$ K $<T_c<2\times
10^7$ K is likely due to the radiative-conductive transition point
lying below the outer H-rich layer, a location where the cooling model
has a thick layer of helium whereas our envelope is pure carbon.
The sharp change in the dotted line at
$L<10^{-4}L_\odot$ is due to surface convection extending down into
the conductive regions, commonly termed ``convective coupling''
\citep{FontBras01}.  To avoid this complication, we have restricted
ourselves to $L>10^{-4}L_\odot$, where the outer convective layer need
not be modelled directly to determine the $L$-$T_c$ relation.

\subsection{Parameter Dependences of White Dwarf Luminosities}

We now discuss how the $L$-$T_c$ relation changes under active
accretion. It is only for low $\timav$'s and high $T_c$'s that the
luminosity flowing through the envelope is dominated by the hot core.
The $T_c$ below which the energy release due to accretion becomes
important is determined by $\timav$; higher values release energy
faster via compressional heating and therefore their luminosity
becomes dominated by accretion at a higher $T_c$.  Thus in Figure
\ref{fig:L-Tc}, while there is
a large area of the curve at $\timav = 10^{-11}M_\odot$ yr$^{-1}$ for
which core cooling dominates at this $M_{\rm acc}$, there is little
such area at $\timav = 10^{-9.4}M_\odot$ yr$^{-1}$.  The upturn in the
luminosity at $T_c\gtrsim3\times 10^7$ K is due to energy released
from fusion of $^3$He, a species not present in abundance in the
cooling models. The onset of the nuclear energy source is quite
sensitive to $M_{\rm acc}$ since the conditions ($T$, $\rho$) at the
base of the H/He layer dominate this effect (see Section \ref{sec:ign}
for details).

\begin{figure}
\plotone{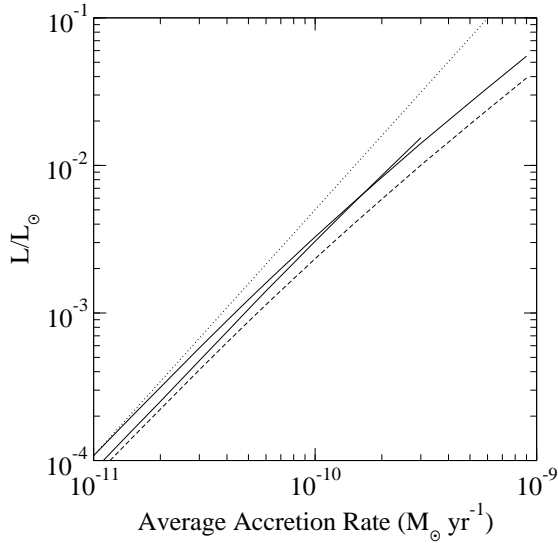}
\caption{
\label{fig:L_0}
Surface luminosity as a function of $\timav$ for very cold WD cores.
The upper solid line is for $M=0.6M_\odot$ and the lower solid line is
for $M=1.0M_\odot$ both with $M_{\rm acc}=5\times10^{-5}M_\odot$.
The dashed line is the values for the top solid line scaled by
$(1.0/0.6)^{-2/3}$ in order to demonstrate that our analytic WD
mass dependence is close to correct. The
dotted line is the analytical approximation discussed in the text. }
\end{figure}

The other major feature of the $L$-$T_c$ curve is the minimum $L$ for
a given $\timav$. In this situation the envelope determines its own
temperature and the surface luminosity becomes independent of
$T_c$. There is a temperature inversion in the envelope: the
temperature rises inward through the radiative layer, flattens out
when conduction becomes dominant, passes through a $T_{\rm max}$ and
then decreases towards the core. The energy liberated by compression
goes into the core as well as coming out the surface.  This lower
limit, $L(T_c\rightarrow 0)$, is shown in Figure \ref{fig:L_0} as a function of
$\timav$. The upper solid curve is for $M=0.6M_\odot$ and $M_{\rm
acc}=5\times 10^{-5}M_\odot$.  Raising $M_{\rm acc}$ does not change
the relation -- a curve for $M_{\rm acc}=10^{-4}M_\odot$ is
indistinguishable -- except to reduce the maximum $L$ at which such a
self-supported envelope state can exist.  A higher $M_{\rm acc}$ leads
to a higher density at the base of the accreted layer and causes
fusion to become important at lower envelope temperatures and
luminosities. 

 The dotted line is a rough approximation found by identifying $T_{\rm
max}$ with the core temperature of a cooling WD, giving $L\propto
T_{\rm max}^{2.5}M$, and by assuming that those losses are matched by
compressional heating, so that $L\approx3\timav kT_{\rm max}/\mu m_p$
(from Appendix \ref{app:estimate}).
Putting these together and using the point at $\timav=10^{-11}M_\odot$
yr$^{-1}$ to set the constant gives $L(\timav)=1.08\times
10^{-4}L_\odot(\timav/10^{-11}M_\odot\ {\rm
yr}^{-1})^{5/3}(M/0.6M_\odot)^{-2/3}$ (dotted line).
Calibrating the cooling luminosity separately from the $\timav=0$
curve on the earlier Figure
\ref{fig:L-Tc} gives a prefactor which differs from this by less than
a factor of 2, demonstrating that the estimate from Appendix
\ref{app:estimate} is good for low $\timav$'s.
While the $L\propto \timav^{5/3}$ dependence is fairly
well matched at low $\timav$'s, the full calculation gives
increasingly lower luminosities at higher $\timav$'s because when
fusion provides some of the energy, $T_{\rm max}$ need not be as high,
so that the luminosity is less.  The lower solid line gives
$L(T_c\rightarrow 0)$ for $M=1.0M_\odot$ and is fairly consistent with
$(1.0/0.6)^{-2/3}$ times the $M=0.6M_\odot$ curve, shown as the dashed
line.  The $M=1.0 M_\odot$ curves follows $L\propto \timav^{5/3}$
to higher $\timav$ than that for $M=0.6M_\odot$.

\begin{figure}
\plotone{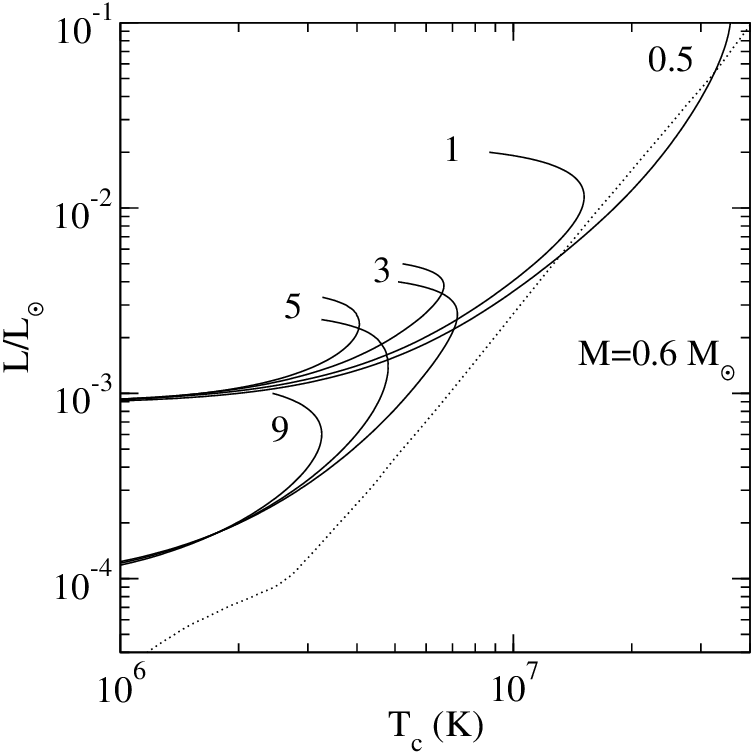}
\caption{
\label{fig:L-Tc_varMacc}
The WD surface luminosity $L$ as a function of $T_c$ for various accumulated layer
masses.  The upper set of solid lines all have $\timav=10^{-10.4}M_\odot$
yr$^{-1}$  and have $M_{\rm acc}= 0.5$, 1, 3, and $5\times10^{-4}M_\odot$ as
indicated.  The lower set of solid lines has $\timav=10^{-11}M_\odot$
yr$^{-1}$ and $M_{\rm acc}= 3,$ 5, and $9\times 10^{-4}M_\odot$.
The WD mass is $0.6M_\odot$ for all cases. The dotted line is the $L$-$T_c$
relation of an isolated cooling DA WD with pure layers of mass $M_{\rm
H}=6\times10^{-5}M_\odot$ and $M_{\rm He}=6\times10^{-3}M_\odot$
\citep{Hans99}.
}
\end{figure}

Finally we explore the dependence of the $L$-$T_c$ relation on $M_{\rm
acc}$.  Generally, the energy released by compression in the envelope
should increase with $M_{\rm acc}$ due to the lower $\mu$ for H/He
relative to C/O as it appears in the compressional heating term (see
eq.~\ref{dLdP}).  Energy from hydrogen fusion should also increase
with $M_{\rm acc}$.  The upper curves in
Figure \ref{fig:L-Tc_varMacc} shows $L$ for
$M_{\rm acc}$ of 0.5, 1, 3, and $5\times 10^{-4}M_\odot$ for a
$0.6M_\odot$ WD accreting at $\timav =10^{-10.4}M_\odot$ yr$^{-1}$.
This accretion rate, $\approx 4\times 10^{-10}M_\odot$ yr$^{-1}$, is
characteristic of gravitational wave driven accretion for dwarf novae
below the period gap \citep{KolbBara99}.
Also shown for comparison is $\timav = 10^{-11}M_\odot$ yr$^{-1}$.
As $M_{\rm acc}$ is increased from 0.5 to 1$\times 10^{-4}M_\odot$,
three important features are apparent.
At $T_c\lesssim 3\times 10^6$ K the
exiting surface luminosity $L$ is nearly unchanged because the additional
energy released in the envelope
is transported into the core with a negligible
rise in $T_{\rm max}$.
For $3\times 10^6$ K$\lesssim
T_c\lesssim 10^7$ K, $T_c$ is high enough that
the additional compressional heating energy is forced outward and acts as
an additive factor on top of the cooling luminosity and there is a
small increase across this range.
At
$T_c\gtrsim 1.5\times 10^7$ K, the density has become high enough that
the large amount of energy released by fusion has destabilized the
envelope at these temperatures.  This is discussed more in Section
\ref{sec:ign}.

\begin{figure}
\plotone{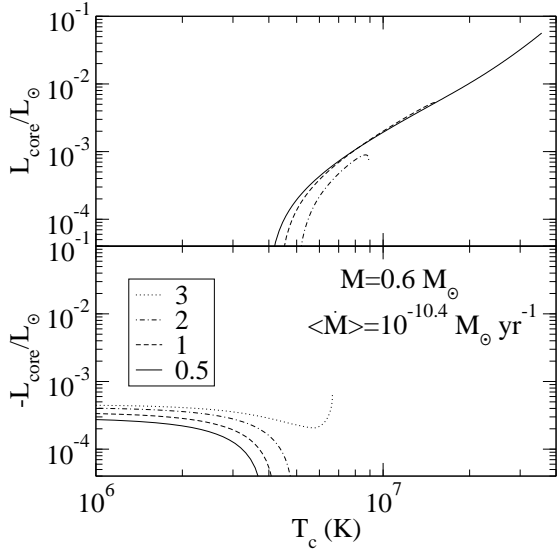}
\caption{
\label{fig:Lin-Tc}
Luminosity, $L_{\rm core}$, entering the accreted layer from the core
below as a function of $T_c$ for
various accumulated layer masses $M_{\rm acc}$ on a $0.6M_\odot$ WD.
The top panel shows the rate of core cooling ($L_{\rm core}$) and the
bottom the rate of heating ($-L_{\rm core}$).  The accretion rate is
set at $\timav=10^{-10.4}M_\odot$ yr$^{-1}$, and $M_{\rm acc}$
takes values of 0.5, 1, 3, and 5$\times 10^{-4}M_\odot$ as indicated.
For clarity the curves have been truncated at the turning point
seen at high $T_c$ in Figure \ref{fig:L-Tc_varMacc}.
}
\end{figure}

The crossover behavior in $T_c$
between core heating at low $T_c$ and cooling at higher $T_c$ can be
seen directly in Figure \ref{fig:Lin-Tc} which shows $L_{\rm core}$,
the luminosity exiting the core into the base of the accreted layer.
In order to maintain a logarithmic scale, two panes are displayed, the
upper displays positive $L_{\rm core}$ and therefore core cooling,
and the lower $-L_{\rm core}$ to show core heating.  When
increasing $M_{\rm acc}$ from 0.5 to $1\times 10^{-4}M_\odot$ (solid and dashed
curves respectively) the core heating increases at
$T_c\lesssim 4\times 10^6$ K. The core cooling, present at higher
temperatures, remains fixed with this $M_{\rm acc}$ increase; 
the additional compressional heating is increasing $L$.
Note that at a single $T_c\simeq 4$-$5\times 10^6$ K, both core-heating and
core-cooling envelope states are available, depending on the value of
$M_{\rm acc}$.

Our survey of parameters is now complete, and in Section
\ref{sec:equil} we use $L_{\rm core}$ to discuss the thermal evolution
of the C/O core.  Of the parameters discussed, $\timav$ will be an
independent variable while
$T_c$ will be found from $\timav$ and $M$. However, $M_{\rm acc}$ has
a maximum value set by the stability of the envelope, which we now
discuss.

\section{Calculating The Ignition Mass}
\label{sec:ign}

The accreted mass, $M_{\rm acc}$, increases until a thermonuclear
instability causes a classical nova explosion at $M_{\rm acc}=M_{\rm
ign}$, which depends principally on the temperature in the accreted
layer.  There are two ignition modes: (1) A traditional thin shell
instability created by the temperature sensitivity of thermonuclear
reactions, or (2) a global instability of the envelope relevant when
$\kappa_{\rm cond} \ll \kappa_{\rm f-f}$ ($T\lesssim 8\times10^6$ K
here).  The mode giving the lower $M_{\rm ign}$ is the active mode,
and each is elaborated below.

\begin{figure}
\plotone{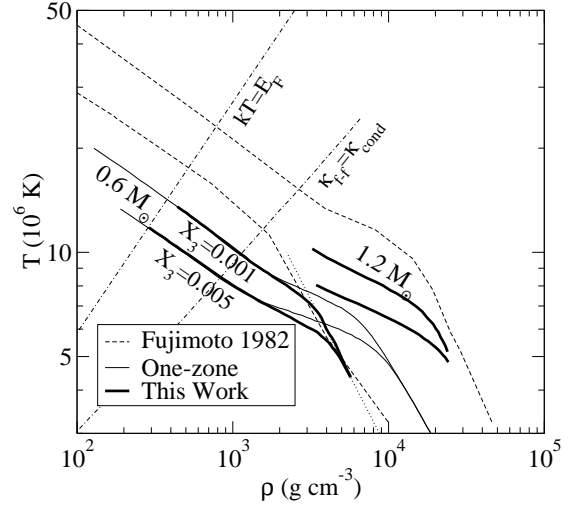}
\caption{
\label{fig:Ign}
Conditions at the base of the H/He accreted layer at ignition of a
classical nova for $M=0.6M_\odot$ and $M=1.2M_\odot$.  The thin solid
curves are the local condition given by equation  (\ref{eq:local}), the
dotted curves are the global ignition condition given by
$L_\odot(T_b/10^8{\ \rm K})^{2.5}(M/0.6M_\odot)= M_{\rm
ign}\epsilon_N(\rho_b,T_b)$, and the thick curves are the actual
conditions found for equilibrated models.  The dashed lines are the
base conditions from ignition calculations by \citet{Fuji82}.
}
\end{figure}

The instability manifests in the $L$-$T_c$ relation as a
turning point, or a maximum $T_c$ as a function of $L$, as seen
earlier in
Figure \ref{fig:L-Tc_varMacc}.  A general discussion of the
relationship between turning points and stability in thin shell
accretion appears in \citet{Pacz83}.
Due to the large heat capacity of the core and the choice of the lower
boundary as described in Section \ref{sec:ourenv},
$T_c$ is essentially constant while $M_{\rm acc}$
increases and the turning point moves to lower temperatures.  If we
define $M_{\rm ign}(T_c)$ to be the $M_{\rm acc}$ at which the turning
point happens at $T_c$, the envelope structures with $M_{\rm
acc}<M_{\rm ign}(T_c)$ form a connected quasi-static sequence at
constant $T_c$ terminating at $M_{\rm ign}$.  Evolution forward in
time from this envelope state must be followed in a fully time
dependent fashion rather than quasi-statically, and so we consider
this terminus the ignition point of the classical nova explosion.
The envelope is assumed to have switched to a ``high'', burning state
about which will make two assumptions: (1) it is short-lived and
therefore of small thermal importance to the core and
(2) during it, most or all of the accreted material is ejected.
The
thick solid curves in Figure \ref{fig:Ign} show the conditions at the
base of the accreted layer at ignition for the models equilibrated in
the manner discussed in \S\ref{sec:equil} for two values of $M$.  Two
values of $X_{^3\rm He}$ are shown at each $M$, representing the range
expected in cataclysmic variable binaries.

The lower dot-dashed line in Figure \ref{fig:Ign} gives the divison
between radiative and conductive heat transport, where $\kappa_{\rm
f-f} = \kappa_{\rm cond}$.  Above this line radiation dominates the
heat transfer at the base of the accreted layer and the instability is
well predicted by a traditional thin shell instability criterion.
Since the scale height of the envelope $h=P/\rho g\ll R$, it has a
positive heat capacity and is therefore subject to a thermal
instability which occurs when the inequality
\begin{equation}
\label{eq:local}
\left(\frac{\partial \epsilon_{\rm cool}}{\partial T}\right)_P
<\left(\frac{\partial \epsilon_{\rm nuc}}{\partial T}\right)_P
\end{equation}
is satisfied at the base of the accreted material.  Here
$\epsilon_{\rm nuc}$ is the energy generated by the fast component of
nuclear burning; the important chains are $^{12}$C$(p,\gamma)^{13}$N
(the first step of CNO), $^1$H$(p,\beta^+\nu)$D$(p,\gamma)^3$He (the
first two steps of the pp chains), $^3$He$(^3$He$,2p)^4$He, and
$^3$He$(\alpha,\gamma)^7$Be$(e^-,\nu)^7$Li$(p,\alpha)^4$He
\citep{Clay83}. To represent local cooling due to heat transfer we
define
\begin{equation}
\epsilon_{\rm cool} \equiv \frac{g^2}{P^2}\frac{4acT^4}{3\kappa}\ ,
\end{equation}
reflecting the rate at which a thermal perturbation on the size of the
pressure scale height, $h$, damps out.  The use of division by the
scale height in lieu of a derivative to form $\epsilon_{\rm cool}$ is
commonly termed a one-zone approximation. See \citet{Pacz83} for an
in-depth example.  Equation (\ref{eq:local}) further approximates that the
pressure response to a small temperature perturbation is
negligible, a good approximation when $h\ll R$.  This ignition
condition is identical to that used by \citet{Fujietal81} to study
X-ray bursts, though our notation is somewhat different.  When the
inequality of equation  (\ref{eq:local}) is satisfied, thermal
perturbations will grow since there is not sufficient cooling to halt
the nuclear runaway.  We evaluate the derivatives using our full
equation of state and $\kappa$ to obtain a maximum temperature at a
given density for the base of H/He layer, shown as the thin solid
curves in Figure \ref{fig:Ign} for the two values of $X_{^3\rm He}$.
When the base temperature in the H/He layer exceeds this value,
ignition occurs.  This agrees with the location of the turning point
in the $L$-$T_c$ relations in the radiative regime (See Figure
\ref{fig:L-Tc_varMacc}).

In the radiative regime the opacity is due to free-free and at
moderate temperatures the triggering nuclear reaction is $^3{\rm
He}+^3{\rm He}$ fusion.  Both the heating and the cooling can be
well approximated by power-laws, so that the ignition condition can be
written in closed form.  Expanding $\epsilon_{\rm nuc}$ about
$T=10^7$ K and using an ideal gas equation of state, equation
(\ref{eq:local}) becomes
\begin{equation}
\rho_{\rm ign}
\gtrsim 1.2\times 10^{3}\ {\rm g\ cm^{-3}}
T_7^{-3.21}
\left(\frac{X_3}{0.001}\right)^{-1/2}
g_8^{1/2}\ ,
\end{equation}
where $T_7 = T/10^7$ K and $g_8 = g/10^8$ cm s$^{-2}$,
matching the radiative portion of curves shown in Figure
\ref{fig:Ign}.

When the base of the envelope becomes degenerate, i.e. when
$\kappa_{\rm cond} \ll \kappa_{\rm f-f}$,
the turning point happens at a lower $M_{\rm acc}$ than the one-zone
model predicts, as shown by the divergence of the thick lines in
Figure \ref{fig:Ign} from the thin lines of the one-zone
prediction.  The second ignition mode has become important and involves
an interaction between the radiative and conductive regions in the
envelope.
In order to clarify the envelope state in this global ignition mode, we
now present how this maximum $T_c$ arises from a simple analytical
model when the base of the accreted layer is in the conductive region.
Define $T_b$ as the temperature at the base of the accreted layer; the
location of the point of transition to conduction at the bottom of the
radiative layer leads to $L=L_0 (T_b/T_0)^{2.5}$, where $L_0$ and
$T_0$ are fiducial parameters. Using a coefficient $K_b$ to
characterize the thermal conduction, the heat coming from the core is
$L_{\rm core} = K_b(T_c-T_b)/z_{bc}$ where $z_{bc}$ is the distance
below the base of the accreted layer at which $T=T_c$ and we ignore
the dependence of $K$ on temperature and density.  If the nuclear
heating during accumulation is approximated by
$L_N=\epsilon_N(\rho_b,T_b)M_{\rm acc}$, where $\rho_b$ is the density
at the base of the accreted layer derived from $P_b=gM_{\rm ign}/4\pi
R^2$, these quantities are related by $L=L_{\rm core}+L_N$, neglecting
compressional heating, or
\begin{eqnarray}
\nonumber
\label{eq:Tcmax} T_c&\simeq&
T_0\left(\frac{L}{L_0}\right)^{0.4}
\\&&
\mbox{}+\frac{z_{bc}}{K}\left[ L - M_{\rm acc}
\left.\epsilon_N\right|_{7\times 10^6{\rm K}} \left(\frac{T_0}{7\times
10^6{\rm K}}\right)^{5.2} \left(\frac{L}{L_0}\right)^{2.1} \right]\ .
\end{eqnarray}
Here we have taken $T_b=T_0 (L/L_0)^{1/2.5}$ and expanded $\epsilon_N$
about $T=7\times 10^{6}$ K
so as to quantify the temperature
dependence of the burning rate in the regime of interest;
only the $p+p\rightarrow ^3$He reaction chain is included at
this low temperature.  For low
luminosities, $T_c$ increases with $L$, but eventually the nuclear
heating becomes strong enough that a maximum $T_c$ is reached.  With
the outer temperature boundary of the conductive region set by the
radiative-conductive transition and the inner temperature boundary
fixed by the thermal inertia of the core, the heat released by nuclear
processes cannot be transported out of the conductive region, and
therefore builds up rapidly, leading to an explosion.

A simple approximation of the maximum
of equation (\ref{eq:Tcmax}) is obtained by finding when the opposing
terms, $L$ and $M_{\rm acc}\epsilon_N$, are comparable.  This gives the
simple form $L_0(T_b/T_0)^{2.5} (M/0.6M_\odot)=M_{\rm
ign}\epsilon_N(\rho_b,T_b)$.  This expresses that $M_{\rm ign}$ is the
accumulated mass for which the outgoing luminosity is wholly provided by
nuclear burning, giving a maximal envelope state before CN runaway.
By expanding $\epsilon_N$ at the lowest $T$ point of the solid line
for $M=0.6M_\odot$, $T=4.36\times 10^6$ K, $\rho = 5.7\times 10^3$ gr
cm$^{-3}$, using only the pp reaction as above but including the
screening correction, we obtain
$\epsilon_N\propto T^{5.73}\rho^{1.17}$.
The
dotted line shown in Figure \ref{fig:Ign} is this form with the prefactor
adjusted to fit the lowest temperature point of our actual envelope
calculation for $M=0.6M_\odot$ (the lower thick solid lines).  Using an
$L$-$T_b$ relation fitted to the $\timav=0$ line in Figure \ref{fig:L-Tc}
gives a result that is parallel and only a factor of 2 lower in density,
mostly due to the neglecting the inward-bound heat from the dropped term.

The dashed lines in Figure \ref{fig:Ign} give the results found by
\citet{Fuji82} for $M=0.6M_\odot$ (lower line) and $1.0M_\odot$ (upper
line).
Our results are consistent with these in the demonstration of two
ignition modes. The largest difference is due to our inclusion of
$^3$He in the accreted matter as a nuclear energy source.  Our
calculation of ignition with $X_3=0$ is roughly 5\% higher in
temperature than \citet{Fuji82} in the radiation-transport dominated
regime.  Much of this 5\% discrepancy is likely from our updated
opacities.

\section{The Thermal Equilibrium of the White Dwarf Core} 
\label{sec:equil}

\begin{figure}
\plotone{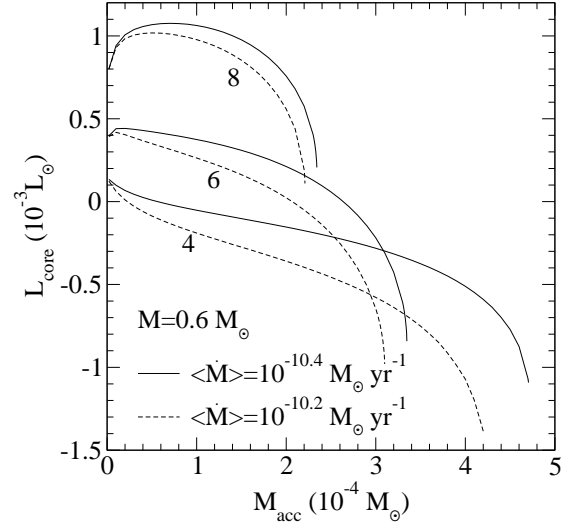}
\caption{
\label{fig:Lcore-Macc}
Luminosity exiting the WD core as a function of $M_{\rm acc}$ for
several values of $T_c$ as indicated in units of $10^6$ K.  All curves
have $M =0.6M_\odot$ and the solid curves have
$\timav=10^{-10.4}M_\odot$ yr$^{-1}$ while the dashed ones
$\timav=10^{-10.2}M_\odot$ yr$^{-1}$.}
\end{figure}

 Section \ref{sec:accenv} introduced the envelope models which provide
$L(M_{\rm acc},\timav,T_c,M)$. These same models also provide $L_{\rm
core}(M_{\rm acc},\timav,T_c,M)$, the luminosity passing through the
base of the envelope from the core.
The variation of $L_{\rm core}$ with
$M_{\rm acc}$ at several $T_c$'s and two accretion rates is shown in
Figure \ref{fig:Lcore-Macc}. Each of these curves extend in $M_{\rm
acc}$ to the CN ignition\footnote{The CN explosion has a negligible
effect on the thermal content of the core due to its short duration
compared to the envelope accumulation time.} and make clear the amount
of heat exchanged between the envelope and the WD core over the CN
cycle, with core cooling occurring at low $M_{\rm acc}$ and core
heating at high $M_{\rm acc}$. This allows us to calculate an
equilibrium core temperature, $T_{\rm c,eq}$, where the heating and
cooling of the core are equal over the CN cycle (Townsley \& Bildsten
2002). At the onset of accretion in the binary, the WD core has
$T_c<T_{\rm c,eq}$ and is heated during the CN cycle. We now find
the time it takes to reach $T_{\rm c,eq}$ at a given $\timav$.

\begin{figure}
\plotone{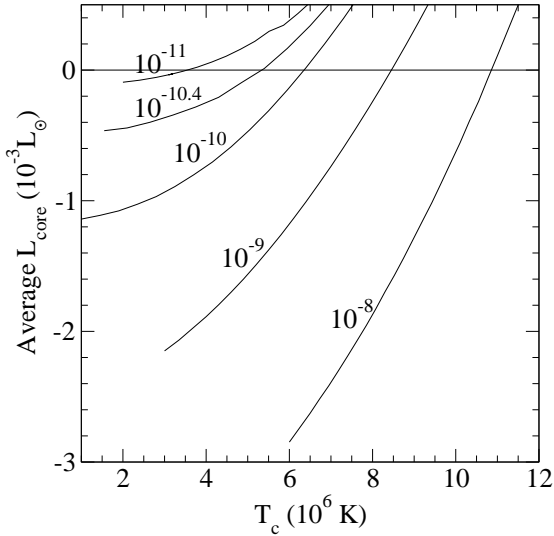}
\caption{
\label{fig:aveL-Tc}
Time averaged 
luminosity escaping from the core during the accretion of matter up
to a classical nova event. The curves have $\timav$ as indicated in units of
$M_\odot$ yr$^{-1}$ and $M_{\rm WD}=0.6M_\odot$.
} \end{figure}

 At low $\timav$ any long-term compression (expansion) of the WD core
due to mass gain (loss) has little impact on the envelope (see Appendix
A). Hence, for simplicity, we assume that the accumulated mass is
ejected in a CN, keeping $M$ constant. We begin by defining an
average core heating rate presuming that $M$ and $T_c$ are constant
while $M_{\rm acc}$ changes, 
\begin{equation}
\label{eq:aveL}
\langle L_{\rm core}\rangle(\timav,T_c,M) =
\frac{1}{t_{CN}}\int _0^{t_{CN}} L_{\rm core}(M_{\rm acc}(t),\timav,T_c,M)dt\ .
\end{equation}
Using this $\langle L_{\rm core}\rangle$ to make statements about
the WD's evolution implicitly assumes that the star's only memory of
the CN cycle is the net heat gained or lost during it, otherwise it
returns to an identical state to that at the same stage in the
previous cycle.
Figure \ref{fig:aveL-Tc} shows the dependence of $\langle L_{\rm
core}\rangle$ on $T_c$ for several accretion rates.  We find that
$\langle L_{\rm core}\rangle$ increases with $T_c$ from two effects:
1) The cooling luminosity is greater at higher $T_c$, causing stronger
core cooling for small $M_{\rm acc}$, and 2) the maximum $M_{\rm acc}$
is larger at lower $T_c$, allowing for a longer period of core heating
when $M_{\rm acc}$ is large. The equilibrium point where $\langle
L_{\rm core}\rangle=0$ for $M=0.6M_\odot$ and
$\timav=10^{-10.4}M_\odot$ yr$^{-1}$, is $T_c\approx5.5\times 10^6$ K.
For $T_c\approx T_{\rm c,eq}$, the envelope transitions from a state
which allows the core to cool at low $M_{\rm acc}$ to a state which
heats the core at high $M_{\rm acc}$. The positive slope of $\langle
L_{\rm core}\rangle$ at $T_c\approx T_{\rm c,eq}$ implies that the
equilibrium state is stable, i.e. a higher $T_c$ at the same $\timav$
causes the core to radiate and lower its temperature.  The figure also
makes clear that $T_{\rm c,eq}$ increases with $\timav$ in order to
maintain equilibrium under the additional heat deposition. 

The specific heat of the liquid WD core is $\approx 3k/\mu_i m_p$,
where $\mu_i\approx 14$ is the ion mean molecular weight (see Appendix
A). When $T_c< T_{\rm c,eq}$, the core is heated by the envelope at
the rate shown in Figure \ref{fig:aveL-Tc},
allowing for a simple integration of
\begin{equation}
\label{eq:evol}
\langle L_{\rm core}\rangle \approx -{3kM\over \mu_i m_p}{{ dT_{\rm c}\over dt}},  
\end{equation}
to yield the time to reach the equilibrium.   For a comparison which
does not depend on the initial $T_c$, we approximate $\langle L_{\rm
core}\rangle \approx (T_c-T_{c,\rm eq})(dL/dT_c)_0$ where the
derivative is evaluated at the equilibrium point, giving the form $T_c
- T_{c,\rm eq} = (T_{\rm in} - T_{c,\rm eq}) e^{-t/\tau}$ where $\tau
= 3kM/\mu_im_p(dL/dT_c)_0$.  This $e$-folding time is 0.22, 0.32, 0.45
and 1.9 Gyr for $\timav = 10^{-8}$, $10^{-9}$, $10^{-10}$, and
$10^{-11}M_\odot$ yr$^{-1}$ respectively.  If $\timav$ changes on a
timescale shorter than this, $T_c$ will not be maintained at its
equilibrium value.  In previous work \citep{BildTown03} we estimated
$(dL/dT_c)_0\approx 3k\timav/2\mu_e m_p$.  This estimate matches the
results from the full envelope calculations at $\timav \sim $
few$\times 10^{-11}M_\odot$ yr$^{-1}$, but the actual $\timav$
dependence is much weaker than estimated due to complications
introduced by the varying ignition mechanism and nuclear energy
source.  For example, heating of the core is relatively less efficient
at $\timav\gtrsim 2\times10^{-10}M_\odot$ yr$^{-1}$ when the base of
the accreted layer never becomes conductive.

The reheating time at a constant $\timav$ from a reasonable starting
temperature can be calculated directly from equation (\ref{eq:evol})
and the curves in Figure \ref{fig:aveL-Tc} for all but the highest
accretion rates.  The time for the core of a $0.6M_\odot$ C/O WD to
heat up from $2.5\times 10^{6}$ K, corresponding to a cooling age 4
Gyr \citep{Salaetal00}, to 90\% of the equilibrium temperature is 0.7
Gyr, 0.9 Gyr and 2.4 Gyr for $10^{-9}$, $10^{-10}$ and
$10^{-11}M_\odot$ yr$^{-1}$ respectively.  The mass that must be
transferred in the binary in each of these cases is thus 0.7, 0.09,
and $0.024M_\odot$.  The latter values are quite reasonable expected
fluxes in terms of evolution of cataclysmic variables (see Howell,
Nelson, \& Rappaport 2001), and thus WDs in CVs with $1.3<P_{\rm orb}<
2$ hours are expected to be in equilibrium.  However, the flux
necessary at $10^{-9}M_\odot$ yr$^{-1}$ is quite high, so that systems
with $P_{\rm orb}> 3$ hours are unlikely to have reached equilibrium.
This will be mitigated to some degree by the fact that all of the
$\sim 0.4M_\odot$ transferred above the period gap occurs at $\timav >
10^{-9}M_\odot$ yr$^{-1}$ \citep{Howeetal01}, which has stronger core
heating.  A system which comes into contact at $P_{\rm orb}\approx 6$
hours should be coming approximately into equilibrium when $P_{\rm
orb}\simeq 3$ hours, just above the period gap.  We defer a detailed
calculation and discussion relating the evolution of the interacting
binary and the thermal state of the WD to a companion paper
\citep{TownBild03}.

\section{Classical Novae Ignition Masses}
\label{sec:res}

  The principal outcome of our calculations is the equilibrium
temperature at which the heating and cooling of the WD core are
balanced over the CN cycle. Once this equilibrium has been reached,
observable quantities do not depend on the age of the WD, but rather
on $M$ and $\timav$, which are determined by the evolution of the
binary system.

\begin{figure}
\plotone{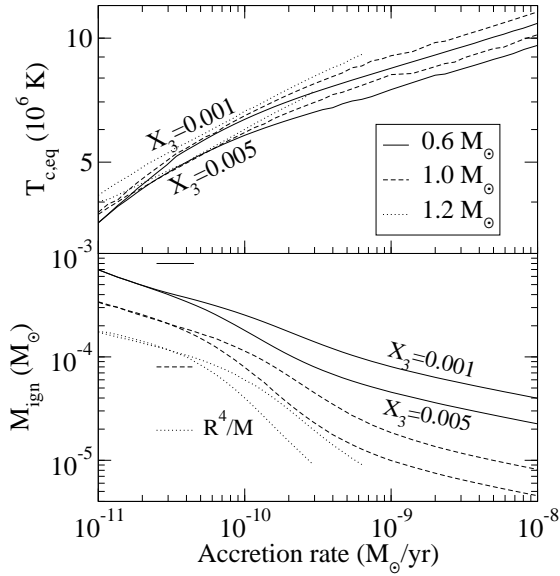}
\caption{
\label{fig:TcMmax}
Equlibrium core temperature and mass of the accreted layer at classical nova
ignition, $M_{\rm ign}$ as a function of time averaged accretion rate,
$\timav$.  Curves are shown for three WD masses, $M$, and two different mass
fractions of $^3$He at each $M$.
}
\end{figure}

The top panel of Figure \ref{fig:TcMmax} shows $T_{c,\rm eq}$ for a
range of $\timav$.  Curves for three values of $M$ are shown, two
for each with $X_{^3\rm He}=0.005$ and 0.001, the upper and lower limits on
expected values for the accreted material  (e.g.\
\citealt{DAntMazz82}; \citealt{IbenTutu84}). The dependence of $T_{c,\rm
eq}$ on these parameters is complicated due to the fact that two
transitions are taking place over this broad range of accretion rates.
There are three regimes: (1) At $\timav \gtrsim 3\times 10^{-10}M_\odot$
yr$^{-1}$, the thermonuclear runaway is well described by a one-zone
model, and the dominant nuclear fuel triggering the explosion is
$^3$He, $T_{c,\rm eq}$ increases with $\timav$ in order to balance the
higher rate of heat deposition in the envelope.  Larger $X_{^3\rm He}$
causes the CN explosion to happen sooner, thus truncating the heating
phase relative to lower $X_{^3\rm He}$ and leading to a lower
$T_{c,\rm eq}$.  The lower limit of this regime depends on $M$ since
higher gravity provides a more stable envelope in the one-zone model.
(2) At $3\times 10^{-11}\lesssim\timav/(M_\odot$ yr$^{-1})\lesssim
3\times10^{-10}$ the one-zone model of ignition is no longer
sufficient (see Section \ref{sec:ign}), but the dominant nuclear heat
source is still $^3$He.  The contrast in $T_{c,\rm eq}$ with $M$ has
become quite small, while the dependence on $X_{^3\rm He}$ remains.
(3) At $\timav\lesssim3\times 10^{-11}M_\odot$ yr$^{-1}$, the dependence
on $X_{^3\rm He}$ is mostly gone and the dominant nuclear heating
source is pp.

While $T_c$ is not directly observable, the resulting ignition masses,
$M_{\rm ign}$, can be compared to measured ejection masses. The bottom
panel of Figure \ref{fig:TcMmax} shows $M_{\rm ign}$ for the
equilibrated stars. This is the first self-consistent evaluation of
the CN ignition mass since we use our calculated $T_{c,\rm
eq}(\timav)$. The ignition mass  decreases as the mass of the WD
increases.  This dependence is familiar from previous work, but is
usually stated as the contrast expected for a constant ignition
pressure, $P_{\rm ign}\approx GMM_{\rm ign}/4\pi R^4$, yielding
$M_{\rm ign}\propto R^4/M$. Clearly the ignition pressure for a fixed
mass is not constant, as $T_c$ changes with $\timav$, thus the simple
scaling does not hold.

The contrast in nuclear energy contributions is clearly demonstrated;
$^{3}{\rm He}$ triggering has a pronounced effect on $M_{\rm
ign}$ at $\timav > 10^{-10}M_\odot$ yr$^{-1}$ and barely any for
$\timav <4\times 10^{-11}M_\odot$ yr$^{-1}$ where pp is the dominant
contributer. Our $M_{\rm ign}$ values with $X_{^3\rm He}=0.001$
are 1/2 to 1/4 of
those found by \citet{Fuji82} (see his Figure 7), the fraction
decreasing with increasing $\timav$. This 
difference is surprisingly small given that his $\timav$'s were
order of magnitude estimates, in contrast to our 
self-consistent calculations. 
The grid of CN simulation results in \citet{PriaKove95} includes
values of $M_{\rm ign}$ (their $m_{\rm acc}$) for $M=0.65$, $1.0$ and
$1.25M_\odot$, $\timav = 10^{-10}$, $10^{-9}$ and $10^{-8}M_\odot$
yr$^{-1}$ with $T_c=10^7$ K.  For $M=0.6$ and $1.0M_\odot$
at $10^{-8}$ and $10^{-9}M_\odot$ yr$^{-1}$,
we find about 1/2 of their $M_{\rm ign}$ for $X_{^3\rm He}=0.001$.
Whereas at $10^{-10}M_\odot$ yr$^{-1}$, again with $X_{^3\rm
He}=0.001$, we find 1, 1.4, and 3.1 times their $M_{\rm ign}$ at $M=
0.6$, 1.0, and $1.2M_\odot$ respectively.  
We understand these differences as
follows. At $\timav= 10^{-10}M_\odot$ yr$^{-1}$ our equilibrium
temperature is less than $10^7$ K thus leading to larger ignition
masses.  Second, our inclusion of $^3$He in the accreted material
decreases $M_{\rm ign}$ with respect to calculations which include
only CNO enrichment leading to our lower $M_{\rm ign}$ values at the
higher accretion rates where $T_{c,\rm eq}$ is close to $10^7$ K, and
somewhat counteracting the affect of the lower $T_c$ at
$\timav=10^{-10}M_\odot\ {\rm yr^{-1}}$.

Our calculations of $M_{\rm ign}$ agree quite well with those of
\citet{MacD84} for $\timav \lesssim 10^{-10.5}M_\odot$ yr$^{-1}$.
This is where pp burning dominates and the base of the accreted layer
is degenerate at ignition.  However, his $T_{c, \rm eq}$'s were not
reported.  At larger $\timav\sim 10^{-9}$-$10^{-8}M_\odot$ yr$^{-1}$
our $M_{\rm ign}$ are nearly a factor of 10 lower than MacDonald.
This discrepancy is larger than can be explained by $^3$He ignition.
As discussed in \citet{MacD83}, the ``cold'' models utilized in
\citet{MacD84} leave out one of three ignition conditions. The omitted
condition turns out to be quite similar to the one-zone condition and
therefore essential at these higher $\timav$'s.

\begin{figure}
\plotone{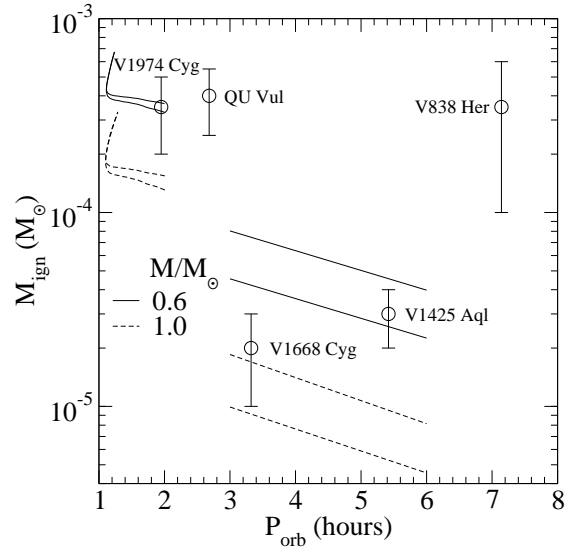}
\caption{
\label{fig:Mign-Porb}
Comparison of our predicted classical nova ignition masses ($M_{\rm
ign}$) to measurements of classical nova ejected masses for systems
which have a measured orbital period in Warner (2002). The ejected
mass references are:
V1974 Cyg (Gehrz et al. 1998),
QU Vul (Shin et al. 1998),
V1668 Cyg (Gehrz et al. 1998),
V1425 Aql (Lyke et al. 2001),
V838 Her (Gehrz et al. 1998).
The
two lines at each $M$ as indicated are for $X_{^3\rm He}=0.001$ and
$0.005$, giving larger and smaller $M_{\rm ign}$ respectively.  At low
periods $\timav$ is that expected from gravitational radiation
\citep{KolbBara99} and at $P_{\rm orb}=3$ and 6 hours we use $\timav
=10^{-9}$ and $10^{-8}M_\odot$ yr$^{-1}$ respectively, which is
approximately that expected for above the period gap due to magnetic
braking \citep{Howeetal01}.
}
\end{figure}

Accumulated masses are not directly measurable, but the mass ejected
in a CN, $M_{\rm ej}$, is known for many systems.  A robust comparison
between $M_{\rm ign}$ and $M_{\rm ej}$ would tell us whether the WD
mass is increasing or decreasing on long timescales during accretion.
Because $M_{\rm ign}$ depends on $\timav$, we must estimate $\timav$
for a given CN system, requiring us to know the orbital period and to
make some presumption about its relationship to $\timav$.  For $P_{\rm
orb}<2$ hours, $\timav$ is that given by angular momentum loss due to
gravitational radiation \citep{KolbBara99}.  For $P_{\rm orb}>3$ hours
mass transfer is driven by magnetic braking.  To approximate the
standard model as presented in e.g.\ \citet{Howeetal01}, we have used two
values, $\timav =10^{-9}M_\odot$ yr$^{-1}$ at $P_{\rm orb}=3$ hours
and $\timav = 10^{-8}M_\odot$ yr$^{-1}$ at $P_{\rm orb} = 6$ hours.
A cross-comparison between $M_{\rm ign}$ and $M_{\rm ej}$ is shown in
Figure \ref{fig:Mign-Porb}, which shows $M_{\rm ign}$ for $M=0.6$ and
$1.0M_\odot$ and $X_{^3\rm He}=0.001$ and $0.005$.  The five $M_{\rm
ej}$ points, which have large errors,
are those systems where both the orbital period and
ejected mass are known.
One of these systems, V1974 Cyg, has a kinematic WD mass estimate of
$0.75-1.07M_\odot$ \citep{Rettetal97}. Other estimates are less certain.
Due to the small number of points,
large errors, and lack of measured WD masses,
the comparison to the current data-set is not
definitive, but does roughly agree, pointing out that 
most of the accreted mass is ejected during the CN. We
have not found overwhelming evidence for excavation of the WD.

There are now over 50 CN with measured orbital periods (Warner 2002).
Our ignition masses will allow for a calculation of the
CN orbital period distribution when combined with CV evolution.
For example, the tenfold contrast in ignition mass above and below
the period gap makes the CN occurence rate nearly a factor of 100
lower below the period gap than above. Currently 10\% of the CN are
below the period gap, if these systems all suffer the same selection
effects then the implied population  below the period gap is 10 times
the number of systems above the period gap.

\begin{figure}
\plotone{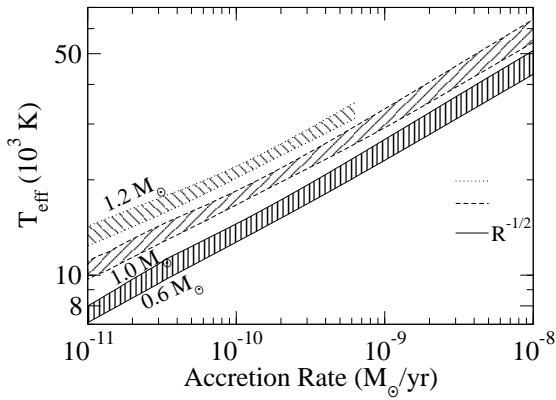}
\caption{
\label{fig:Teff}
Ranges of WD $T_{\rm eff}$ as a function of
 accretion rate for WD mass $M=0.6$, 1.0, and
$1.2M_\odot$ (bottom to top).  The minimum is evaluated when $M_{\rm
acc}=0.05M_{\rm ign}$ and the maximum when $M_{\rm acc} = 0.95 M_{\rm
ign}$.  For comparison, in 1 Gyr a $0.6(1.0)M_\odot$ isolated WD cools
to $T_{\rm eff} = 9\times 10^3 (14\times 10^3)$ K \citep{Salaetal00}.
}
\end{figure}

Our envelope models also predict $T_{\rm eff}=(L/4\pi
R^2\sigma_{SB})^{1/4}$ which increases with $M_{\rm acc}$.  Figure
\ref{fig:Teff} shows, as a function of $\timav$ for $T_c=T_{c,\rm
eq}$,
the range of $T_{\rm eff}$ traversed between $M_{\rm acc} = 0.05M_{\rm
ign}$ and $0.95M_{\rm ign}$ for $M=0.6$, $1.0$, and $1.2M_\odot$.  We
find that the surface flux $L$ does not depend strongly on mass, and
thus roughly $T_{\rm eff}\propto R^{-1/2}$.  Short horizontal lines
with this dependence are shown for comparison.  The WD $T_{\rm eff}$
has been measured for a number of Dwarf Novae systems when the disk is
in quiescence \citep{Sion99}, and a detailed comparison to such
measurements is being published separately \citep{TownBild03}.

\section{Conclusion and Discussion}
\label{sec:concl}

A quasi-static WD envelope model has been used to find the equilibrium
core temperature, $T_{c,\rm eq}$ of an accreting WD under the
approximation that exactly the accreted layer is ejected in each
classical nova outburst.  Such an equilibrium can be reached in
$\lesssim1$ Gyr, allowing for CVs to come close to equilibrium by the
time they reach $P_{\rm orb}\simeq 3$ hours, and insuring
equilibration for $P_{\rm orb} < 2$ hours.  Having solved for
$T_{c,\rm eq}$ it is then possible to predict the CN ignition mass
$M_{\rm ign}$ and the thermal contribution to the WD $T_{\rm eff}$.
This is the first determination of $M_{\rm ign}$ which utilizes a WD
state determined from $\timav$ and the cooling and heating of the core
during the CN buildup, making it fully self consistent.
When compared with several papers from the
CN literature (\citealt{Fuji82}; \citealt{PriaKove95};
\citealt{MacD84}) we find that the inclusion of
$^3$He leads to lower $M_{\rm ign}$ for $\timav \gtrsim
10^{-10}M_\odot\ {\rm yr^{-1}}$, and that for $\timav$ below this the
particular author's assumption concerning $T_c$, which we calculate
consistently, is a determining factor.
When compared
with observed ejected masses, our $M_{\rm ign}$ are compatible with
$M_{\rm ej}\simeq M_{\rm ign}$.  Had this comparison favored either
$M_{\rm ej} > M_{\rm ign}$ or the opposite, the equilibrium would take
place at some $T_c$ such that $L_{\rm core}$ has the value appropriate
for the implied secular change in core mass, which can be
calculated using the formalism presented in Appendix \ref{app:core}.

Over twenty Dwarf Novae have been observed in quiescence, when the
accretion rate is low and the WD photosphere is detected and $T_{\rm
eff}$ measured.  The theoretical work presented here is compared to these
observations in \citet{TownBild03}, allowing us to constrain the WD
mass and the time averaged accretion rate, $\timav$. We show there,
for systems with $P_{\rm orb} \lesssim 2$ hours, that if $\timav$ is
given by gravitational radiation losses alone, then the WD masses
are $>0.8M_\odot$.  An alternative conclusion is that the masses are
closer to $0.6M_\odot$ and $\timav$ is 3-4 times larger than that
expected from gravitational radiation losses.

It is well known that an isolated WD will pulsate when its $T_{\rm
eff}$ is in the approximate range 11000-12000 K \citep{Bergetal95}.
While a difference in the atmospheric composition, H/He mixture
under accretion versus pure hydrogen on the non-accreting case, will
shift this range, it is likely that a similar pulsation mechanism will
be active in accreting WDs.  Our calculations indicate that accreting
WDs with $M=0.6$-$1.0M_\odot$ should be near this range when $\timav
=$ few$ \times 10^{-11}M_\odot$ yr$^{-1}$ (See Figure \ref{fig:Teff}).
This $\timav$ is typical of that expected in CVs when accretion is driven
by emission of gravitational radiation, $P_{\rm orb}< 2$ hours
\citep{KolbBara99}.  In fact
one system, GW Lib, has been found which does exhibit precisely this
type of variability \citep{VanZetal00,Szkoetal02}.
Using the interior models
developed here, we are now undertaking a seismological study of these
systems.  This offers the tantalizing possibility of facilitating
measurement of the WD mass and spin and determination of gross
internal structural features such as the size of the accreted layer.

 We thank Ed Sion for numerous conversations, Paula Szkody for 
up to date information on the DN observations, Brad Hansen for
insights on modern WD modeling, Francesca D'Antona for questioning
our equilibria, and the referee for constructive criticism.
This research was supported by the
National Science Foundation under Grants 
PHY99-07949 and AST02-05956. Support for this work was provided by
NASA through grant AR-09517.01-A from STScI, which is operated by
AURA, Inc., under NASA contract NAS5-26555. D. T. is an NSF Graduate
Fellow and L. B. is a Cottrell Scholar of the Research Corporation.

\appendix

\section{The Thermal State of the Underlying White Dwarf}
\label{app:core}

  The thermal state of the WD core can affect the envelope structure
through the heat flow into or out of it on the evolutionary times we
are considering. Due to varying ejected mass in classical novae
outbursts, we do not actually know whether the WD core is increasing
or decreasing in mass. However, we show here that for a C/O core, the
compression or decompression rate of the WD core will have little
impact on the overall envelope calculations. This is less clear for
those WDs with pure helium cores that are likely present in binaries
below the period gap (e.g.\ \citealt{Howeetal01}). Our
discussion follows closely the insights from
\citet{NomoSugi77}, \citet{Nomo82}, and \citet{Hernetal88}.

  The WDs of interest here consist mostly of $^{12}{\rm C}$ and
$^{16}{\rm O}$ so we construct the deep interior (far beneath the
freshly accreted hydrogen and helium) using the degenerate electron
equation of state for $\mu_e=2$, neglecting the small $^{22}{\rm Ne}$
fraction. The ions completely dominate the thermodynamics and thermal
conductivity of the liquid WD interior. For a classical one component
plasma (OCP) with ion separation, $a$, defined by $a^3=3/4\pi n_i$,
where $n_i=\rho/A m_p$, the importance of Coulomb physics for the ions
is measured by
\begin{equation}
\label{eq:gamma}
\Gamma\equiv {(Ze)^2\over a kT} =57.7\rho_6^{1/3}
\left(10^7 \ {\rm K}\over T\right) 
\left(Z\over
8\right)^2\left(16\over A\right)^{1/3},
\end{equation}
where $\rho_6=\rho/10^6 \ {\rm gr \ cm^{-3}}$. We will work in the
liquid regime, $\Gamma< 175$ (\citealt{PoteChab00} and references
therein). For a $0.6(1.2)M_\odot$ pure Oxygen WD, this requires that
the core temperature, $T_c$, exceeds $5(16)\times 10^{6} \ {\rm K}$
(this is within the range expected for the low mass WDs from section
5). Following the recent work of \citet{DeWietal96}
and \citet{ChabPote98}, we used the linear sum rule to
evaluate the Coulomb energy in the mixed liquid phase and the
resulting specific heat, $c_V$. We then integrated throughout the
stellar model to find the total heat capacity of the white dwarf,
$C_V=\int c_V dM$, as a function of the isothermal WD temperature. In
agreement with \citet{PoteChab00}, we find that $C_V$ is
nearly that of a classical crystal, $C_V=3k(M/\mu_i m_p)$, for
temperatures in the range of the melting value and slightly above and
below.

\begin{figure}
\plotone{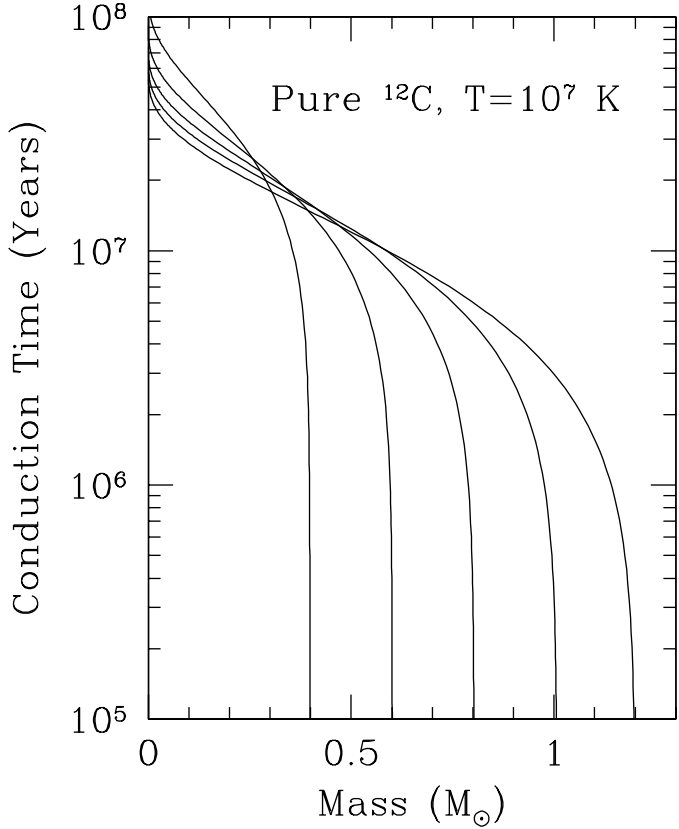}
\caption{The thermal conduction time ($t_{\rm cond}$ in equation
[\ref{eq:tcond}]) from the exterior of a pure carbon WD to an interior
mass point. The curves are for isothermal WDs ($T=10^7$ K) 
with masses $M=0.4,0.6,0.8,1.0$ and $1.2M_\odot$. 
\label{fig:tcond}}
\end{figure}

 Electron conduction dominates the heat transport in the deep liquid
WD interior and we use the work of \citet{YakoUrpi80} for our
estimates of the thermal conductivity, $K$, when electron-ion
collisions dominate.  For an
isothermal WD, the time it takes for conduction to transport heat
from the outside edge, $R$, to an interior radius $r$ is
\begin{equation}
\label{eq:tcond} 
t_{\rm cond}={1\over 4}\left(\int_r^R \left(c_V\over \rho K\right)^{1/2} \rho
dr\right)^2,
\end{equation}
as shown by \citet{HenyLEcu69}. This is just a slightly more
sophisticated version of what \citet{NomoSugi77} call
$\tau_h=c_V \rho H_p^2/K$, the time for heat transport across a
pressure scale height, $H_p$. Since $c_V$ is nearly independent of
temperature, the main $T$ scaling of $t_{\rm cond}$ is from the
conductivity, yielding $t_{\rm cond}\propto 1/T$. Figure
\ref{fig:tcond} plots $t_{\rm cond}$ as a function of the mass
location within pure, isothermal ($T=10^7$ K) carbon WDs of masses
$M=0.4,0.6,0.8,1.0$ and 1.2 $M_\odot$. The timescale is only slightly
different for pure oxygen WD's.

   The compression of the WD interior is adiabatic if accretion
compresses it on a timescale much shorter than $t_{\rm cond}\approx
(2-4)\times 10^{7}(10^7 {\rm K}/T)\ {\rm yr}$. To relate the
compression rate to $\dot M$, we start at the WD core and write
$d\ln\rho_c/dt=(d\ln\rho_c/d\ln M)d\ln M/dt$ (e.g. \citealt{Nomo82}). For
a low-mass WD, $\alpha=d\ln\rho_c/d\ln M=2$, but $\alpha$ increases as
M approaches the Chandrasekhar mass, having values of $\alpha=3.27,
4.26, 6.17,$ and $ 10.71 $ for $M=0.6, 0.8, 1.0,$ and
$1.2M_\odot$. For example, the accretion rate required to
adiabatically compress the interior of a $0.8M_\odot$ WD when $T=10^7
{\rm K}$ is $> 6\times 10^{-9} M_\odot \ {\rm yr^{-1}}$. Simulations
for rates this rapid have been carried out (e.g. Nomoto \& Iben 1985)
and clearly show the central adiabatic compression. Since the ions
dominate the entropy there, we calculate the adiabat in the liquid
state using the internal energy given by \citet{ChabPote98}
to find
\begin{equation} 
\left({d\ln T\over d\ln \rho}\right)_{\rm ad}\approx {0.91+0.14\Gamma^{1/3}\over 1.22+0.41
\Gamma^{1/3}}, 
\end{equation} 
within 1\% of the adiabat given by \citet{Hernetal88} for $\Gamma>1$
to crystallization. For
accretion rates where the compression time and conduction time are
comparable, there is clearly time for heat to be transported, but not
adequate time for maintaining an isothermal core.

However, for $\dot M \ll 6\times 10^{-9} M_\odot \ \rm{yr^{-1}}$, there
is time for the core to remain isothermal under the compression
(Hernanz et al. 1988), simplifying the treatment of the WD core for
the $\dot M$'s of most interest to us. Unlike our treatment of the
atmosphere, for the deep interior, we need to account for the changing
global structure of the WD. Nomoto (1982) showed that this is greatly
simplified by rewriting the entropy equation in terms of the variable
$q=M_r/M$, where $M_r$ is the mass interior to $r$ and $M$ is the
total mass. Under accretion, we relate $\dot q/q=\dot M_r/M_r-\dot
M/M$, and rewrite the time derivative as $d/dt=\partial/\partial
t|_q+\dot q \partial / \partial q|_t$.  In the entropy equation, we
want the Lagrangian time derivative $d/dt|_{M_r}$, so that $\dot
q=-q\dot M/M$, and the entropy equation simply becomes
\begin{equation} 
T\left(\partial s\over \partial t\right)_q-T{d\ln M\over dt}\left(\partial s\over
\partial {\rm ln}q\right)_t=-{\partial L_r\over \partial M_r}, 
\end{equation} 
under the action of compression. Integrating this equation tells us
the luminosity exiting the surface of the WD core
\begin{equation}
\label{eq:lumins}
L={d\ln M\over d\ln t}\int_0^M T \left(\partial s\over \partial \ln
q\right)_t dM_r-\int_0^M T\left(\partial s\over \partial t\right)_q
dM_r, 
\end{equation}
as clearly written by Nomoto \& Sugimoto (1977). 

The advantage to
writing the terms in this manner becomes apparent once we use 
\begin{equation}
\label{eq:ds} 
ds ={k\over \mu_i m_p}\left[a{\rm
dlnT}-b{\rm dln \rho}\right],
\end{equation} 
where $a=1.22+0.41\Gamma^{1/3}$, and $b=0.91+0.14\Gamma^{1/3}$ are
presumed to be roughly constant throughout the WD. If the ions were an
ideal gas, $a=3/2$ and $b=1$. Presuming that the
WD core is always isothermal at $T=T_c$, then
equation (\ref{eq:lumins}) becomes
\begin{equation} 
L{\mu_i m_p\over kT_c}=-{d\ln M\over dt}\int_0^M b\left(d\ln\rho\over
 d\ln q\right)_t dM_r -aM{d\ln T_c\over dt}+\int_0^M
b\left(d\ln\rho\over d\ln M\right)_q{d\ln M\over dt}dM_r.
\end{equation}
We rewrite this more clearly by regrouping terms to obtain
\begin{equation}
\label{eq:lumins2}
L+{aM k\over \mu_i m_p}{d T_c\over dt}=
-{bkT_c\over \mu_i m_p}{d\ln M\over dt}\int_0^M \left[\left(d\ln\rho\over
 d\ln q\right)_t - \left(d\ln\rho\over d\ln M\right)_q\right]dM_r.
\end{equation} 
 The term $d\ln\rho/d\ln M|_q$ is the generalization of the piece
discussed earlier in the context of the central point. It gives the
compression rate at a fixed mass coordinate, $q$, due to accretion.
Nomoto (1982) showed that it is independent of $q$
within the deep interior, so that  $\alpha=d\ln\rho_c/
d\ln M\approx d\ln\rho/d\ln M|_q$, and equation (\ref{eq:lumins2})
becomes 
\begin{equation}
L+{aM k\over \mu_i m_p}{d T_c\over dt}=
{\alpha bkT_c\over \mu_i m_p}{d M\over dt}
-{bkT_c\over \mu_i m_p}{d\ln M\over dt}\int_0^M M_r d\ln \rho,
\end{equation} 
where the last term gets large near the surface and is evaluated in
Appendix B as part of the outer envelope integration (see equation
B3). At $\Gamma\approx 80$, $a=3$ and $b=1.5$, so for a $M=0.8
M_\odot$ WD we get the simple relation for compression of the deep
interior of the WD
\begin{equation}
L+{aM k\over \mu_i m_p}{d T_c\over dt}=
{6.4kT_c\over \mu_i m_p}{d M\over dt},
\end{equation} 
implying that the core can remain at a fixed temperature during
compression from accretion as long as the exiting luminosity 
is
\begin{equation}
\label{eq:lcore}
L_{\rm core}={6.4kT_c\over \mu_i m_p}{d M\over dt}. 
\end{equation} 
Comparison of this with the corresponding estimate for the accreted
layers from Appendix \ref{app:estimate}, equation
(\ref{eq:lestimate}), demonstrates that, even if the WD core is
accumulating mass on a secular timescale (i.e. no ejection of matter),
then, since $\mu_i\approx 14$, such an exiting luminosity as $L_{\rm
core}$ would be a small perturbation on the thermal state of the
envelope. This statement would change for a helium WD, where
$\mu_i=4$, in which case the core luminosity could modify the envelope
structure.

\section{Analytical Envelope Integration}
\label{app:estimate}

Here we present the integration of equation (\ref{dLdP}) from Section
\ref{sec:accenv} using a simple model for a
WD consisting of a radiative outer envelope and a degenerate deeper layer
that is isothermal.  Dropping the nuclear burning and the time derivative
term from equation (\ref{eq:heateq}) leads to the integral
\begin{equation}
L= \timav \int_0^P T\frac{ds}{dP}dP\ .
\end{equation}
To allow a simple integration, in the envelope we make the
approximation that the luminosity is constant with depth and that the
opacity has the form $\kappa\propto \rho T^{-3.5}$ so that $P^2\propto
T^{8.5}$.  The entropy of an ideal gas is $s=k\ln(T^{3/2}/\rho)/\mu
m_p$, which can be expressed in terms of pressure alone, allowing
evaluation of $ds/dP$, and
\begin{equation}
L_{\rm env} = 0.41\frac{k}{\mu m_p} \timav \int T\frac{dP}{P}
=1.75\timav\frac{kT_c}{\mu m_p}\ .
\end{equation}
Where $P^2\propto T^{8.5}$ has been used to evaluate the integral, the
term from the outer boundary has been dropped, and the pressure at the inner
boundary (the transition to degeneracy) has been written in terms of $T_c$.

The degenerate section is nearly isothermal and the entropy is completely in
the ions, so that $s=k\ln(T^{3/2}/\rho)/\mu_i m_p$, and the density is related
to the pressure by $P\propto \rho^{5/3}$ from the surrounding
non-relativistic degenerate electrons.  From this $ds/dP$ can be evaluated,
but some limits must be chosen.  The lower limit is the pressure at which the
transition to degeneracy occurs, this is simply $P_{\rm tr}=T_c^{5/2}$ in cgs
units for a solar mixture.  The upper limit is the pressure at the base of
the accreted envelope $P_{\rm base}=gM_{\rm acc}/4\pi R^2$.  From our
results,
at $\timav = 10^{-10}M_\odot$ yr$^{-1}$, $T_c=8\times 10^6$ K
and $M_{\rm ign}= 3\times 10^{-4}M_\odot$.  Using these values gives
$P_{\rm tr}=1.8\times 10^{17}$ erg cm$^{-3}$ and $P_{\rm
base}=6.5\times 10^{18}$ erg cm$^{-3}$ and thus
\begin{equation}
L_{\rm deg} = \frac{3}{5}\frac{kT_c}{\mu_i m_p}\timav \int d\ln P
=2.15\timav \frac{kT_c}{\mu_i m_p}\ .
\end{equation}
Although the limits taken actually depend on $T_c$, they appear in a
logarithm so only the strongest temperature dependence is shown explicitly.
Using $\mu/\mu_i = 0.6/1.3$ for solar material to eliminate $\mu_i$ and
adding the luminosity from the nondegenerate and degenerate portions of the
accreted layer gives 
\begin{equation}
\label{eq:lestimate}
L\approx\frac{ 3 \timav kT_c}{\mu m_p}\ .
\end{equation}
This fiducial estimate of the compressional heating proves helpful in our
analytic understanding of our initial results.



\end{document}